# Comparative Performance Investigations of different scenarios for 802.15.4 WPAN

**Sukhvinder S. Bamber and Ajay K. Sharma**

**Department of Computer Science & Engineering, National Institute of Technology,
Jalandhar, Punjab 144011, India**

**Abstract**
This paper investigates the performance of WPAN based on various topological scenarios like: cluster, star and ring. The comparative results have been reported for the performance metrics like: Throughput, Traffic sent, Traffic received and Packets dropped. Cluster topology is best in comparison with star and ring topologies as it has been shown that the throughput in case of cluster topology (79.887 kbits / sec) as compared to star (31.815 kbits / sec) and ring (1.179 kbits / sec).

## 1. Introduction

Brought up in 1990s, Wireless Personal Area Networks (WPANs) are the youngest members in the networking hierarchy. WPAN consists of number of nodes distributed in a given area, where a specific phenomenon must be measured and monitored. The construction of WPAN includes the physical deployment and the organization of its logical topology. In this framework, the formation of network topology is a research issue of increasing importance. Recently IEEE approved 802.15.4 standard defining the Medium Access Control sub layer (MAC) and the physical layer (PHY) for low-rate, Wireless Personal Area Networks (LR-WPAN). IEEE 802.15.4 devices typically operate in limited personal operating space.

The objective of this paper is to analyze the performance of various network topologies of IEEE 802.15.4 WPAN. The novelty of the work resides in the evaluation of key performance parameters that can be influenced by the topology. The performance has been analyzed through extensive simulations to capture the behaviour of some key performance parameters. These investigations are usable to configure IEEE 802.15.4 WPAN to select a suitable topology.

The organization of the paper is as follows: Section 1 gives brief introduction of WPAN and objectives of this paper. Section 2 constitutes the system description. Section 3 shows the results and discussions derived form the simulation carried out on different topological scenarios of 802.15.4 WPAN using OPNET® Modeler 14.5. Finally Section 4 concludes the paper.

## 2. System Description

The simulation model implements physical and Media Access Control (MAC) layers defined in IEEE 802.15.4 standard. The OPNET® Modeler is used for developing its sophisticated graphical user interface. This version of simulation model supports point – to – point, star, ring, mesh and cluster topologies where communication is established between devices, called inside the model end devices and a single central controller called PAN coordinator. Each device operating in the network has a unique IP address.

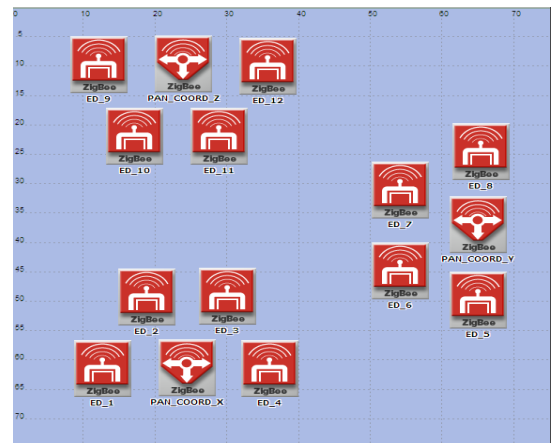

(a)

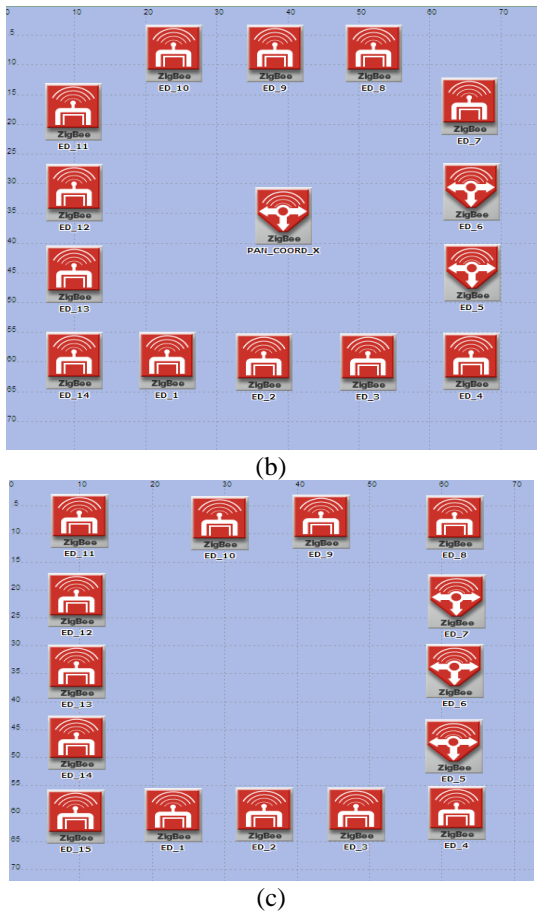

(b)

(c)

Figure: 1 WPAN topologies – (a) Cluster (b) Star (c) Ring

Figure: 1 shows three different topological scenarios of 802.15.4 WPAN: Cluster, Star and Ring. Cluster topology as shown in Figure: 1(a) contains three PAN Coordinators (Fully Functional Devices) which manages their local networks and communicate with each other, rest of the devices in the scenario are end devices that communicate with their respective PAN coordinator in peer to peer mode. Similarly Figure: 1(b) contains only one PAN coordinator and rest of the devices are end devices – all connected to the PAN coordinator but not among themselves. In order to communicate each end device has to communicate to the PAN coordinator first and then the PAN coordinator communicates to the destination end device i.e. no two end devices can directly communicate but only through the PAN coordinator. Figure: 1(c) contains no PAN coordinator but all end devices and each end device communicates to the immediate next device thus forming a ring.

Figure: 1 shows two types of devices: PAN coordinator and end devices having the difference between there parametric values as shown in the table 1 but all have the same type of node model as shown below:

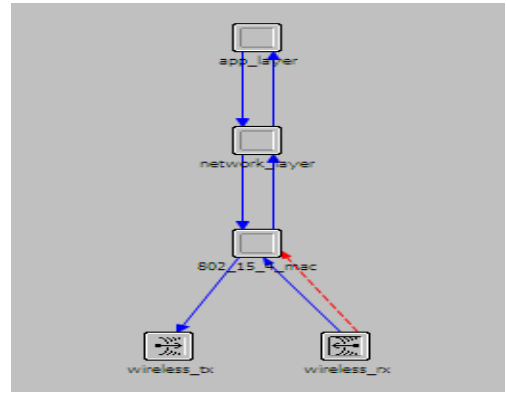

Figure: 2 Node Model for 802.15.4 WPAN

As it can be seen from the figure: 2, node model has four layers: physical, MAC, network and application layers. Physical layer consists of a transmitter and a receiver compliant to the IEEE 802.15.4 specification, operating at 2.4 GHz frequency band and data rate equal to 250 kbps. MAC layer implements slotted CSMA/CA. Network layer is responsible for end to end delivery of packets including routing through intermediate hosts. Finally the topmost Application layer is responsible for generation and reception of traffic.

Corresponding process model for the MAC layer of WPAN that deals with each and every operation on the data that is to be transmitted in the entire scenario is as follows:

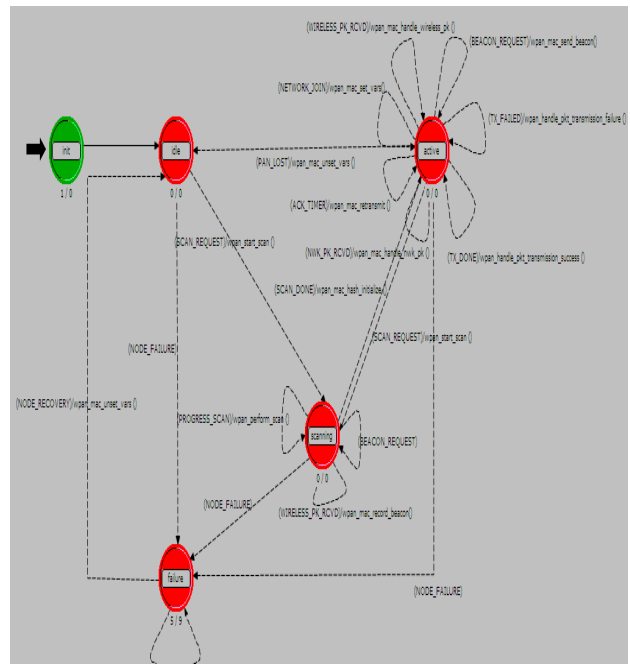

Figure: 3 Process Model for WPAN MAC

Above diagram represents a process model for any device in WPAN. It's functioning is based upon slotted CSMA/CA mechanism, foundation for which is constructed by the various states like: Init - used for initializing MAC, Idle - for introducing delays in order to make the maximum use of the resources, Active – time period during which packets are transmitted, Scanning – for requesting the access to channel on which to transmit the packets.

Here three different topological scenarios have been created namely: Cluster, Ring and Star. Each scenario has different parametric values as shown below:

Table 1: Parametric values for cluster, ring and star topology

| Parameter | Cluster | | Star | | Ring |
|---|---|---|---|---|---|
| | PAN co_ord | End device | PAN co_ord | End device | End device |
| MAC Parameters | | | | | |
| ACK Wait Duration | 0.05 | | | | |
| Number of Retransmissions | 5 | | | | |
| CSMA Parameters | | | | | |
| Minimum Back off Exponent | 3 | | | | |
| Maximum number of Back offs | 4 | | | | |
| Channel Sensing Duration | 0.1 | | | | |
| Application Traffic | | | | | |
| Packet Interarrival Time | constant (1.0) | Exponential (1.0) | constant (1.0) | Exponential (1.0) | Exponential (1.0) |
| Packet Size | Constant (1024) | Exponential (1024) | Constant (1024) | Exponential (1024) | Exponential (1024) |
| Start Time | Uniform (20,21) | Exponential (1) | Uniform (20,21) | Exponential (1) | Exponential (1) |
| Stop Time | infinity | infinity | infinity | infinity | infinity |
| Destination | all coordinators and routers | PAN_Coord | All nodes | PAN_Coord | Immediate Next Node |

## 3. Results and Discussion

Simulation has been carried out for different topologies of WPAN. In this section various results have been presented and discussed to show the impact of different topologies on the performance factors like: throughput, data traffic sent, data traffic received, packets dropped etc.

### 3.1 Throughput

Throughput is the average number of bits or packets successfully received or transmitted by the receiver or transmitter channel per second. Figure: 4 shows the throughput for the cluster, star and ring topologies respectively.

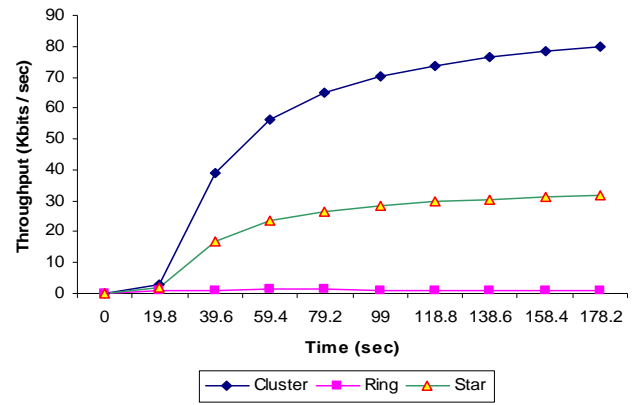

Figure: 4 Throughputs for Cluster, Ring and Star Topologies

Figure: 4 shows that Throughput is 79.887, 31.815 and 1.179 kbits/sec for cluster, star and ring topologies respectively. It has been observed that throughput is maximum in case of cluster topology because cluster topology is communicating on the basis of three fully functional devices called PAN coordinators which are more efficient as compared to the end devices while star topology is having one fully functional device and the ring topology is based only on the end devices. This fully functional device acts as routers, repeaters, amplifiers or regenerators. Also in cluster topology total load of the network is divided among the local PAN coordinators as a result of which lesser collisions and lesser packet drops takes place as a result of which the throughput is maximum in case of cluster topology. It has been observed that throughput is minimum in case of ring topology because in ring topology when the station wants to transmit data has to attain the token so only one system at a time can transmit data which leads to the lowering of throughput of the network.

### 3.2 Packets Dropped

Packets dropped can be defined as the packets that are unable to reach the destination from the source and are lost on the way due to the factors like signal degradation over the network medium, oversaturated network links, corrupted packets, faulty networking hardware, faulty networking drivers etc. Figure: 5 shows the packets dropped for the Cluster, Ring and Star topologies.

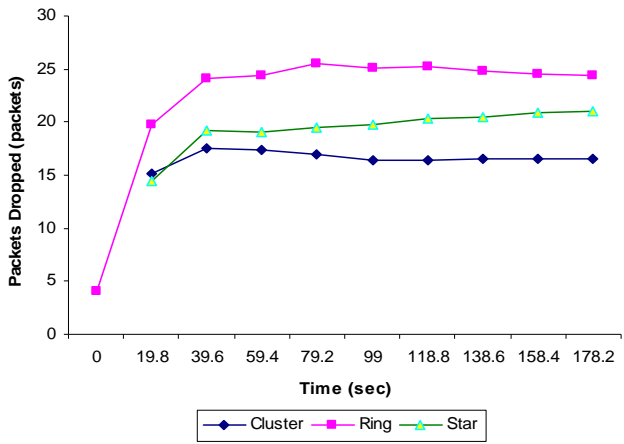

Figure: 5 Packets Dropped for Cluster, Ring and Star Topologies

Figure: 5 reveals that packets dropped are 24.38, 21.08 and 16.56 packets for ring, star and cluster topologies respectively. It has been observed that packets dropped are maximum in case of Ring topology because it works on the basis of token system i.e. only that system is allowed to transmit packets that is having the control of token, after the packet has been transmitted with IP address of the destination, the packets goes from one system to another system till the IP address matches and that particular system receives the packet, this is a bit time consuming process and leads to the increase of traffic in the network as a result of which the packets get lost in the network because of the oversaturated network links, collisions, delays etc.

3.3 Data Traffic Received

Data traffic received can be defined as number of bits of the data received per unit time. Figure: 6 depict the data traffic received for the cluster, star and ring topologies respectively in WPAN.

Figure: 6 shows that the data received by the cluster, star and ring topologies is 276.89, 151.34 and 11.12 kbits/sec respectively. It clearly indicates that the traffic received is maximum in case of cluster topology because cluster topology makes the use of PAN coordinators for communication and these PAN coordinators are responsible for traffic generation and routing as the routing tables are maintained by the PAN coordinators only. Also the lesser collisions, lower packet loss leads to the maximum data traffic in case of cluster topology.

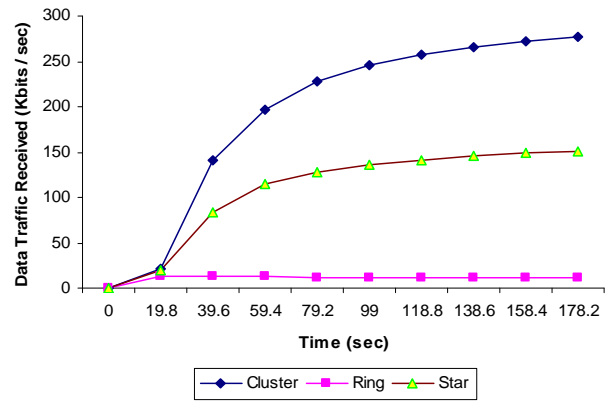

Figure: 6 Data Traffic Received for Cluster, Ring and Star Topologies

Also it is observed that data traffic received is minimum in case of ring topology as it makes use of all end devices and no PAN coordinator also it works on the basis of token system which leads to more collisions and packet losses which ultimately reduces traffic received.

It is also observed that packets dropped are minimum in case of cluster topology as the cluster topology is communicating through the PAN coordinators. Each PAN coordinator manages the data of its networks locally as a result of which lesser collisions take place while communicating. Also being fully functional devices they are better routers, regenerators of the data as compared to the end devices so comparatively lesser dropped packets.

3.4 Data Traffic Sent

Data traffic sent is defined as the total number of data bits sent by the source to the destination per unit time irrespective of the condition whether all of the data bits reach the destination or not. Figure: 7 indicate the data traffic sent for cluster, star and ring topologies respectively in WPAN.

Figure: 7 reveals that the data traffic sent for cluster, star and ring topologies is 47.875, 19.940 and 1.543 kbits/sec respectively. It is clearly observed that data sent is maximum in case of cluster topology because cluster topology makes the use of PAN coordinators for communication and these PAN coordinators are responsible for traffic generation and routing as the routing tables are maintained by the PAN coordinators only. Also the lesser collisions, lower packet loss leads to the maximum data traffic in case of cluster topology.

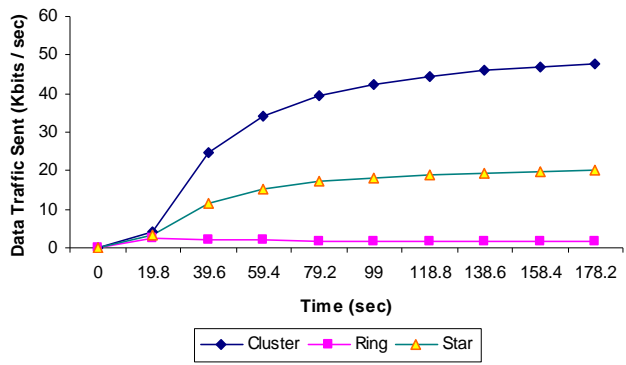

Figure: 7 Data Traffic Sent for Cluster, Ring and Star Topologies

Also it has been observed that data traffic sent is minimum in case of ring topology as it makes use of all end devices and no PAN coordinator, also it works on the basis of token system in which data flows in one direction only which results in the reduction of traffic sent. In addition to that ring topology is extended point – to – point topology suffering from larger collisions and time-outs, both of these factors leads to the reduction of traffic sent.

## 4. Conclusions

This paper presents the performance study of WPAN in three different topological scenarios: cluster, ring and star. The results indicate that throughput is maximum (79.887 kbits/sec) in case of cluster topology while it is 31.815 kbits/sec in star topology and least in case of ring topology i.e. 1.179 kbits/sec. Packets dropped (24.38 packets) are maximum in case of ring topology followed by star (21.08 packets) and least in cluster topology (16.56 packets). Results also conclude that traffic sent (47.875 kbits/sec) and traffic received (276.89 kbits/sec) are maximum in case of cluster topology. From the results and discussions in the section IV it is concluded that cluster topology is more efficient and best suited for the Wireless Personal Area Networks (WPAN). Finally it is concluded that the performance of cluster-based WPAN is best for the application where hardware cost constraints are not significant.

**Sukhvinder S Bamber** completed his B.Tech (Computer Science & Engineering) from PTU, Jalandhar in 2001 and M-Tech (Computer Science & Engineering) from Allahabad Agricultural Institute (Deemed University) in 2007. Currently he is pursuing PhD in the department of Computer science & Engineering at National Institute of Technology, Jalandhar.

**Ajay K Sharma** received his BE in Electronics and Electrical Communication Engineering from Punjab University Chandigarh, India in 1986, MS in Electronics and Control from Birla Institute of Technology (BITS), Pilani in the year 1994 and PhD in Electronics Communication and Computer Engineering in the year 1999. His PhD thesis was on "Studies on Broadband Optical Communication Systems and Networks". From 1986 to 1995 he worked with TTTI, DTE Chandigarh, Indian Railways New Delhi, SLIET Longowal and National Institute of technology (Erstwhile Regional Engineering College), Hamirpur HP at various academic and administrative positions. He has joined National Institute of Technology (Erstwhile Regional Engineering College) Jalandhar as Assistant Professor in the Department of Electronics and Communication Engineering in the year 1996. From November 2001, he has worked as Professor in the ECE department and presently he working as Professor in Computer Science & Engineering in the same institute. His major areas of interest are broadband optical wireless communication systems and networks, dispersion compensation, fiber nonlinearities, optical soliton transmission, WDM systems and networks, Radio-over-Fiber (RoF) and wireless sensor networks and computer communication. He has published 215 research papers in the International/National Journals/Conferences and 12 books. He has supervised 11 Ph.D. and 28 M.Tech theses. He has completed three R&D projects funded by Government of India. He was associated to implement the World Bank project of 209 Million for Technical Education Quality Improvement programme of the institute. He is technical reviewer of reputed international journals. He has been appointed as member of technical Committee on Telecom under International Association of Science and Technology Development (IASTD) Canada for the term 2004-2007 and he is Life member of Indian Society for Technical Education (I.S.T.E.), New Delhi